\PassOptionsToPackage{unicode}{hyperref}
\PassOptionsToPackage{hyphens}{url}
\PassOptionsToPackage{dvipsnames,svgnames,x11names}{xcolor}
\documentclass[
]{article}
\usepackage{amsmath,amssymb}
\usepackage{lmodern}
\usepackage{iftex}
\ifPDFTeX
  \usepackage[T1]{fontenc}
  \usepackage[utf8]{inputenc}
  \usepackage{textcomp} 
\else 
  \usepackage{unicode-math}
  \defaultfontfeatures{Scale=MatchLowercase}
  \defaultfontfeatures[\rmfamily]{Ligatures=TeX,Scale=1}
\fi
\IfFileExists{upquote.sty}{\usepackage{upquote}}{}
\IfFileExists{microtype.sty}{
  \usepackage[]{microtype}
  \UseMicrotypeSet[protrusion]{basicmath} 
}{}
\makeatletter
\@ifundefined{KOMAClassName}{
  \IfFileExists{parskip.sty}{%
    \usepackage{parskip}
  }{
    \setlength{\parindent}{0pt}
    \setlength{\parskip}{6pt plus 2pt minus 1pt}}
}{
  \KOMAoptions{parskip=half}}
\makeatother
\usepackage{xcolor}
\NewDocumentCommand\citeproctext{}{}
\NewDocumentCommand\citeproc{mm}{%
  \begingroup\def\citeproctext{#2}\cite{#1}\endgroup}
\makeatletter
 \let\@cite@ofmt\@firstofone
 \def\@biblabel#1{}
 \def\@cite#1#2{{#1\if@tempswa , #2\fi}}
\makeatother
\newlength{\cslhangindent}
\newlength{\csllabelwidth}
\newenvironment{CSLReferences}[2] 
 {\begin{list}{}{%
  \setlength{\itemindent}{0pt}
  \setlength{\leftmargin}{0pt}
  \setlength{\parsep}{0pt}
  \ifodd #1
   \setlength{\leftmargin}{\cslhangindent}
   \setlength{\itemindent}{-1\cslhangindent}
  \fi
  \setlength{\itemsep}{#2\baselineskip}}}
 {\end{list}}
\usepackage{calc}

\ifLuaTeX
\usepackage[bidi=basic]{babel}
\else
\usepackage[bidi=default]{babel}
\fi
\babelprovide[main,import]{american}

\def\languageshorthands#1{}
\ifLuaTeX
  \usepackage{selnolig}  
\fi
\IfFileExists{bookmark.sty}{\usepackage{bookmark}}{\usepackage{hyperref}}
\IfFileExists{xurl.sty}{\usepackage{xurl}}{} 
\hypersetup{
  pdftitle={Dyad: a binary-star dynamics and statistics library for
Python},
  pdfauthor={Amery Gration},
  pdflang={en-US},
  colorlinks=true,
  linkcolor={Maroon},
  filecolor={Maroon},
  citecolor={Blue},
  urlcolor={Blue},
  pdfcreator={LaTeX via pandoc}}

\title{Dyad: a binary-star dynamics and statistics library for Python}

\definecolor{c53baa1}{RGB}{83,186,161}
\definecolor{c202826}{RGB}{32,40,38}
\def \rorglobalscale {0.1}
\newcommand{\rorlogo}{%
\begin{tikzpicture}[y=1cm, x=1cm, yscale=\rorglobalscale,xscale=\rorglobalscale, every node/.append style={scale=\rorglobalscale}, inner sep=0pt, outer sep=0pt]
  \begin{scope}[even odd rule,line join=round,miter limit=2.0,shift={(-0.025, 0.0216)}]
    \path[fill=c53baa1,nonzero rule,line join=round,miter limit=2.0] (1.8164, 3.012) -- (1.4954, 2.5204) -- (1.1742, 3.012) -- (1.8164, 3.012) -- cycle;
    \path[fill=c53baa1,nonzero rule,line join=round,miter limit=2.0] (3.1594, 3.012) -- (2.8385, 2.5204) -- (2.5172, 3.012) -- (3.1594, 3.012) -- cycle;
    \path[fill=c53baa1,nonzero rule,line join=round,miter limit=2.0] (1.1742, 0.0669) -- (1.4954, 0.5588) -- (1.8164, 0.0669) -- (1.1742, 0.0669) -- cycle;
    \path[fill=c53baa1,nonzero rule,line join=round,miter limit=2.0] (2.5172, 0.0669) -- (2.8385, 0.5588) -- (3.1594, 0.0669) -- (2.5172, 0.0669) -- cycle;
    \path[fill=c202826,nonzero rule,line join=round,miter limit=2.0] (3.8505, 1.4364).. controls (3.9643, 1.4576) and (4.0508, 1.5081) .. (4.1098, 1.5878).. controls (4.169, 1.6674) and (4.1984, 1.7642) .. (4.1984, 1.8777).. controls (4.1984, 1.9719) and (4.182, 2.0503) .. (4.1495, 2.1132).. controls (4.1169, 2.1762) and (4.0727, 2.2262) .. (4.0174, 2.2635).. controls (3.9621, 2.3006) and (3.8976, 2.3273) .. (3.824, 2.3432).. controls (3.7505, 2.359) and (3.6727, 2.367) .. (3.5909, 2.367) -- (2.9676, 2.367) -- (2.9676, 1.8688).. controls (2.9625, 1.8833) and (2.9572, 1.8976) .. (2.9514, 1.9119).. controls (2.9083, 2.0164) and (2.848, 2.1056) .. (2.7705, 2.1791).. controls (2.6929, 2.2527) and (2.6014, 2.3093) .. (2.495, 2.3487).. controls (2.3889, 2.3881) and (2.2728, 2.408) .. (2.1468, 2.408).. controls (2.0209, 2.408) and (1.905, 2.3881) .. (1.7986, 2.3487).. controls (1.6925, 2.3093) and (1.6007, 2.2527) .. (1.5232, 2.1791).. controls (1.4539, 2.1132) and (1.3983, 2.0346) .. (1.3565, 1.9436).. controls (1.3504, 2.009) and (1.3351, 2.0656) .. (1.3105, 2.1132).. controls (1.2779, 2.1762) and (1.2338, 2.2262) .. (1.1785, 2.2635).. controls (1.1232, 2.3006) and (1.0586, 2.3273) .. (0.985, 2.3432).. controls (0.9115, 2.359) and (0.8337, 2.367) .. (0.7519, 2.367) -- (0.1289, 2.367) -- (0.1289, 0.7562) -- (0.4837, 0.7562) -- (0.4837, 1.4002) -- (0.6588, 1.4002) -- (0.9956, 0.7562) -- (1.4211, 0.7562) -- (1.0118, 1.4364).. controls (1.1255, 1.4576) and (1.2121, 1.5081) .. (1.2711, 1.5878).. controls (1.2737, 1.5915) and (1.2761, 1.5954) .. (1.2787, 1.5991).. controls (1.2782, 1.5867) and (1.2779, 1.5743) .. (1.2779, 1.5616).. controls (1.2779, 1.4327) and (1.2996, 1.3158) .. (1.3428, 1.2113).. controls (1.3859, 1.1068) and (1.4462, 1.0176) .. (1.5237, 0.944).. controls (1.601, 0.8705) and (1.6928, 0.8139) .. (1.7992, 0.7744).. controls (1.9053, 0.735) and (2.0214, 0.7152) .. (2.1474, 0.7152).. controls (2.2733, 0.7152) and (2.3892, 0.735) .. (2.4956, 0.7744).. controls (2.6016, 0.8139) and (2.6935, 0.8705) .. (2.771, 0.944).. controls (2.8482, 1.0176) and (2.9086, 1.1068) .. (2.952, 1.2113).. controls (2.9578, 1.2253) and (2.9631, 1.2398) .. (2.9681, 1.2544) -- (2.9681, 0.7562) -- (3.3229, 0.7562) -- (3.3229, 1.4002) -- (3.4981, 1.4002) -- (3.8349, 0.7562) -- (4.2603, 0.7562) -- (3.8505, 1.4364) -- cycle(0.9628, 1.7777).. controls (0.9438, 1.7534) and (0.92, 1.7357) .. (0.8911, 1.7243).. controls (0.8623, 1.7129) and (0.83, 1.706) .. (0.7945, 1.7039).. controls (0.7588, 1.7015) and (0.7252, 1.7005) .. (0.6932, 1.7005) -- (0.4839, 1.7005) -- (0.4839, 2.0667) -- (0.716, 2.0667).. controls (0.7477, 2.0667) and (0.7805, 2.0643) .. (0.8139, 2.0598).. controls (0.8472, 2.0553) and (0.8768, 2.0466) .. (0.9025, 2.0336).. controls (0.9282, 2.0206) and (0.9496, 2.0021) .. (0.9663, 1.9778).. controls (0.9829, 1.9534) and (0.9914, 1.9209) .. (0.9914, 1.8799).. controls (0.9914, 1.8362) and (0.9819, 1.8021) .. (0.9628, 1.7777) -- cycle(2.6125, 1.3533).. controls (2.5889, 1.2904) and (2.5553, 1.2359) .. (2.5112, 1.1896).. controls (2.4672, 1.1433) and (2.4146, 1.1073) .. (2.3529, 1.0814).. controls (2.2916, 1.0554) and (2.2228, 1.0427) .. (2.1471, 1.0427).. controls (2.0712, 1.0427) and (2.0026, 1.0557) .. (1.9412, 1.0814).. controls (1.8799, 1.107) and (1.8272, 1.1433) .. (1.783, 1.1896).. controls (1.7391, 1.2359) and (1.7052, 1.2904) .. (1.6817, 1.3533).. controls (1.6581, 1.4163) and (1.6465, 1.4856) .. (1.6465, 1.5616).. controls (1.6465, 1.6359) and (1.6581, 1.705) .. (1.6817, 1.7687).. controls (1.7052, 1.8325) and (1.7388, 1.8873) .. (1.783, 1.9336).. controls (1.8269, 1.9799) and (1.8796, 2.0159) .. (1.9412, 2.0418).. controls (2.0026, 2.0675) and (2.0712, 2.0804) .. (2.1471, 2.0804).. controls (2.223, 2.0804) and (2.2916, 2.0675) .. (2.3529, 2.0418).. controls (2.4143, 2.0161) and (2.467, 1.9799) .. (2.5112, 1.9336).. controls (2.5551, 1.8873) and (2.5889, 1.8322) .. (2.6125, 1.7687).. controls (2.636, 1.705) and (2.6477, 1.6359) .. (2.6477, 1.5616).. controls (2.6477, 1.4856) and (2.636, 1.4163) .. (2.6125, 1.3533) -- cycle(3.8015, 1.7777).. controls (3.7825, 1.7534) and (3.7587, 1.7357) .. (3.7298, 1.7243).. controls (3.701, 1.7129) and (3.6687, 1.706) .. (3.6333, 1.7039).. controls (3.5975, 1.7015) and (3.5639, 1.7005) .. (3.5319, 1.7005) -- (3.3226, 1.7005) -- (3.3226, 2.0667) -- (3.5547, 2.0667).. controls (3.5864, 2.0667) and (3.6192, 2.0643) .. (3.6526, 2.0598).. controls (3.6859, 2.0553) and (3.7155, 2.0466) .. (3.7412, 2.0336).. controls (3.7669, 2.0206) and (3.7883, 2.0021) .. (3.805, 1.9778).. controls (3.8216, 1.9534) and (3.8301, 1.9209) .. (3.8301, 1.8799).. controls (3.8301, 1.8362) and (3.8206, 1.8021) .. (3.8015, 1.7777) -- cycle;
  \end{scope}
\end{tikzpicture}
}

\usepackage[affil-it]{authblk}
\usepackage{orcidlink}
\author[1%
  ]{Amery Gration%
    \,\orcidlink{0000-0003-1379-6696}\,%
    }

\affil[1]{Astrophysics Research Group, University of Surrey, Guildford,
GU2 7XH, United Kingdom%
    \,\protect\href{https://ror.org/00ks66431}{\protect\rorlogo}\,%
  }
\date{12 February 2026}

\begin{document}
\maketitle

\section{Summary}\label{summary}

Dyad is a Python library for studying the dynamics of binary stars
considered as gravitational two-body systems. By convention,
astrophysicists designate the brighter of the two components of a binary
star as a reference body, which they call the `primary
star\textquotesingle, and the dimmer of the two components as a subject
body, which they call the `secondary star\textquotesingle. In a frame
centred on the primary star, the secondary star then moves on an
elliptical orbit with one focus located at the origin. This orbit can be
specified by its orbital elements, namely the semimajor axis (which
specifies the size of the ellipse), eccentricity (which specifies the
shape of the ellipse), and true anomaly (which specifies the secondary
star\textquotesingle s location on the ellipse) along with the longitude
of its ascending node, its inclination, and its argument of pericentre
(which together specify the orientation of the ellipse).

The dynamics of the system are completely determined by the two
stars\textquotesingle{} masses and the secondary star\textquotesingle s
orbital elements. In a population of binary stars these eight parameters
vary from member to member and can each be treated as a random variable
having some probability distribution. Dyad provides a class,
\texttt{dyad.TwoBody}, and a module, \texttt{dyad.stats}, for dealing
with such a population of binary stars. The \texttt{dyad.TwoBody} class
represents a gravitational two-body system while the \texttt{dyad.stats}
module provides a suite of classes representing the probability
distributions of stellar mass, mass ratio, and orbital elements. Dyad
implements these probability distributions in the same way that SciPy
(\citeproc{ref-virtanen2020}{Virtanen et al., 2020}) implements its
probability distributions (see, for example,
\href{https://docs.scipy.org/doc/scipy/reference/generated/scipy.stats.norm.html}{\texttt{scipy.stats.norm}},
\href{https://docs.scipy.org/doc/scipy/reference/generated/scipy.stats.lognorm.html}{\texttt{scipy.stats.lognorm}},
or
\href{https://docs.scipy.org/doc/scipy/reference/generated/scipy.stats.expon.html}{\texttt{scipy.stats.expon}}).

You can initialize \texttt{dyad.TwoBody} by specifying the primary and
secondary bodies\textquotesingle{} masses together with the secondary
body\textquotesingle s orbital elements and then use that
class\textquotesingle s methods to compute the two
bodies\textquotesingle{} angular momenta, total energies, and
eccentricity vectors as well as positions, velocities, kinetic energies,
and potential energies either in the primary star\textquotesingle s
frame or the centre-of-mass frame. The \texttt{dyad.stats} module
includes (but is not limited to) classes representing the probability
distributions of (1) stellar masses as proposed by Kroupa
(\citeproc{ref-kroupa2001}{2001}) and Salpeter
(\citeproc{ref-salpeter1955}{1955}), and (2) the mass ratios and orbital
elements of binary stars as proposed by Duquennoy \& Mayor
(\citeproc{ref-duquennoy1991}{1991}) and Moe \& Di Stefano
(\citeproc{ref-moe2017}{2017}). You can use it to evaluate the
probability density functions, cumulative distribution functions, and
inverse cumulative distribution functions of these quantities, as well
as to compute their moments, i.e.~their means, variances, skewnesses,
\emph{etc}. Most importantly, you can use \texttt{dyad.stats} to
generate samples of these quantities. By using \texttt{dyad.TwoBody} and
\texttt{dyad.stats} together you can therefore generate a custom
representation of a population of binary stars.

\section{Statement of need}\label{statement-of-need}

I wrote Dyad to perform the work on binary-star population dynamics that
my colleagues and I presented in Gration et al.
(\citeproc{ref-gration2025}{2025}). Galactic dynamicists typically treat
a galaxy as a population of single stars moving in the gravitational
potential generated by those stars together with a halo of dark matter.
They construct a model of its kinematics (say, a model of the
probability density of its stars\textquotesingle{} positions and
velocities) which they then fit to observations (say, those
stars\textquotesingle{} on-sky positions and line-of-sight velocities)
in order to infer the physical properties of that galaxy, for example
its total mass. However, a large number of stars are binary, meaning
that these models are misspecified since the kinematics of stars in a
binary-rich galaxy are different from those in a binary-free galaxy. If
observations contain binary stars then these inferences will be wrong.
The error will be small for large disc galaxies but will be large for
small spheroidal galaxies. For these small spheroidal galaxies the total
mass is always proportional to the velocity dispersion and in our paper
my colleagues and I quantified the error in the inferred galactic mass
by constructing the velocity distribution using Dyad.

The modelling of binary-rich galaxies is an active area of research
(see, for example, the papers by \citeproc{ref-minor2010}{Minor et al.,
2010}; \citeproc{ref-rastello2020}{Rastello et al., 2020}; and
\citeproc{ref-arroyo-polonio2023}{Arroyo-Polonio et al., 2023}).
However, there is no publicly available software dedicated to the field.
The situation is different for the allied field of population synthesis,
in which stellar physicists generate a representation of a population of
binary stars by simulating the dynamical and chemical evolution of some
initial population. There is a large amount of software that can perform
these simulations. Amongst the available packages are COSMIC
(\citeproc{ref-breivik2020}{Breivik et al., 2020}), COMPAS
(\citeproc{ref-riley2022}{Riley et al., 2022}), and binary\_c
(\citeproc{ref-hendriks2023}{Hendriks \& Izzard, 2023}). These packages
invariably allow you to generate the initial population by sampling
stellar mass, mass-ratio, and orbital elements. In that respect they
provide functionality similar to Dyad\textquotesingle s. But typically
they provide only the sampling routine and no further functionality,
such as the ability to evaluate the probability density functions
themselves. Moreover, each package provides a different library of
sampling routines and these do not always allow for correlations between
quantities. None of them was the laboratory that we required for our
work. That laboratory was Dyad, which I hope others will find useful
too.

\section{Acknowledgements}\label{acknowledgements}

This work was supported by UK Research and Innovation grant
MR/S032223/1.

\section*{References}\label{references}
\addcontentsline{toc}{section}{References}

\protect\phantomsection\label{refs}


\begin{CSLReferences}{1}{0}
\bibitem[\citeproctext]{ref-arroyo-polonio2023}
Arroyo-Polonio, J. M., Battaglia, G., Thomas, G. F., Irwin, M. J.,
McConnachie, A. W., \& Tolstoy, E. (2023). Binary star population of the
{Sculptor} dwarf galaxy. \emph{Astronomy \& Astrophysics}, \emph{677},
A95. \url{https://doi.org/10.1051/0004-6361/202346843}

\bibitem[\citeproctext]{ref-breivik2020}
Breivik, K., Coughlin, S., Zevin, M., Rodriguez, C. L., Kremer, K., Ye,
C. S., Andrews, J. J., Kurkowski, M., Digman, M. C., Larson, S. L., \&
Rasio, F. A. (2020). {COSMIC} variance in binary population synthesis.
\emph{The Astrophysical Journal}, \emph{898}, 71.
\url{https://doi.org/10.3847/1538-4357/ab9d85}

\bibitem[\citeproctext]{ref-duquennoy1991}
Duquennoy, A., \& Mayor, M. (1991). Multiplicity among solar-type stars
in the solar neighbourhood II. {Distribution} of the orbital elements in
an unbiased sample. \emph{Astronomy and Astrophysics}, \emph{248}, 485.

\bibitem[\citeproctext]{ref-gration2025}
Gration, A., Hendriks, D. D., Das, P., Heber, D., \& Izzard, R. G.
(2025). Stellar velocity distributions in binary-rich ultrafaint dwarf
galaxies. \emph{Monthly Notices of the Royal Astronomical Society},
\emph{543}, 1120--1132. \url{https://doi.org/10.1093/mnras/staf1481}

\bibitem[\citeproctext]{ref-hendriks2023}
Hendriks, D. D., \& Izzard, R. G. (2023). {binary\_c-Python}: A
{Python-based} stellar population synthesis tool and interface to
binary\_c. \emph{Journal of Open Source Software}, \emph{8}(85), 4642.
\url{https://doi.org/10.21105/joss.04642}

\bibitem[\citeproctext]{ref-kroupa2001}
Kroupa, P. (2001). On the variation of the initial mass function.
\emph{Monthly Notices of the Royal Astronomical Society}, \emph{322}(2),
231--246. \url{https://doi.org/10.1046/j.1365-8711.2001.04022.x}

\bibitem[\citeproctext]{ref-minor2010}
Minor, Q. E., Martinez, G., Bullock, J., Kaplinghat, M., \& Trainor, R.
(2010). Correcting velocity dispersions of dwarf spheroidal galaxies for
binary orbital motion. \emph{The Astrophysical Journal}, \emph{721},
1142--1157. \url{https://doi.org/10.1088/0004-637X/721/2/1142}

\bibitem[\citeproctext]{ref-moe2017}
Moe, M., \& Di Stefano, R. (2017). Mind your {Ps} and {Qs}: The
interrelation between period ({P}) and mass-ratio ({Q}) distributions of
binary stars. \emph{The Astrophysical Journal Supplement Series},
\emph{230}(2), 15. \url{https://doi.org/10.3847/1538-4365/aa6fb6}

\bibitem[\citeproctext]{ref-rastello2020}
Rastello, S., Carraro, G., \& Capuzzo-Dolcetta, R. (2020). Effect of
binarity in star cluster dynamical mass determination. \emph{The
Astrophysical Journal}, \emph{896}, 152.
\url{https://doi.org/10.3847/1538-4357/ab910b}

\bibitem[\citeproctext]{ref-riley2022}
Riley, J., Agrawal, P., Barrett, J. W., Boyett, K. N. K., Broekgaarden,
F. S., Chattopadhyay, D., Gaebel, S. M., Gittins, F., Hirai, R., Howitt,
G., Justham, S., Khandelwal, L., Kummer, F., Lau, M. Y. M., Mandel, I.,
de Mink, S. E., Neijssel, C., Riley, T., van Son, L., \ldots{} Team
Compas. (2022). Rapid stellar and binary population synthesis with
{COMPAS}. \emph{The Astrophysical Journal Supplement Series},
\emph{258}, 34. \url{https://doi.org/10.3847/1538-4365/ac416c}

\bibitem[\citeproctext]{ref-salpeter1955}
Salpeter, E. E. (1955). The luminosity function and stellar evolution.
\emph{The Astrophysical Journal}, \emph{121}, 161--167.
\url{https://doi.org/10.1086/145971}

\bibitem[\citeproctext]{ref-virtanen2020}
Virtanen, P., Gommers, R., Oliphant, T. E., Haberland, M., Reddy, T.,
Cournapeau, D., Burovski, E., Peterson, P., Weckesser, W., Bright, J.,
van der Walt, S. J., Brett, M., Wilson, J., Millman, K. J., Mayorov, N.,
Nelson, A. R. J., Jones, E., Kern, R., Larson, E., \ldots{} van
Mulbregt, P. (2020). {SciPy} 1.0: Fundamental algorithms for scientific
computing in {Python}. \emph{Nature Methods}, \emph{17}(3), 261--272.
\url{https://doi.org/10.1038/s41592-019-0686-2}

\end{CSLReferences}
\end{document}